\documentclass[preprint,amsmath,amssymb]{revtex4}

\usepackage{graphicx}
\usepackage{dcolumn}
\usepackage{bm}
\usepackage{slashbox}

\begin{document}

\title{
Replica-Permutation Method with the Suwa-Todo Algorithm beyond
the Replica-Exchange Method
} 

\author{
Satoru G. Itoh\footnote{Electronic address: itoh@ims.ac.jp} and
Hisashi Okumura\footnote{Electronic address: hokumura@ims.ac.jp}
}
\affiliation{
Department of Theoretical and Computational Molecular Science \\
Institute for Molecular Science \\
Okazaki, Aichi 444-8585, Japan \\
Department of Structural Molecular Science \\
The Graduate University for Advanced Studies \\
Okazaki, Aichi 444-8585, Japan\\
}

\begin{abstract}

We propose a new method for molecular dynamics and Monte Carlo simulations,
which is referred to as the replica-permutation method (RPM),
to realize more efficient sampling than the replica-exchange method (REM).
In RPM not only exchanges between two replicas 
but also permutations among more than two replicas are performed. 
Furthermore, 
instead of the Metropolis algorithm,
the Suwa-Todo algorithm is employed for replica-permutation trials
to minimize its rejection ratio. 
We applied RPM to
particles in a double-well potential energy,
Met-enkephalin in vacuum, and a C-peptide analog of ribonuclease A in explicit water.
For a comparison purposes,
replica-exchange molecular dynamics simulations were also performed.
As a result, RPM sampled not only the temperature space but also
the conformational space more efficiently than REM for all systems.
From our simulations of C-peptide,
we obtained the $\alpha$-helix structure with
salt-bridges between Gly2 and Arg10 which is known in experiments.
Calculating its free-energy landscape, 
the folding pathway was revealed from an extended structure to
the $\alpha$-helix structure with the salt-bridges.
We found that the folding pathway consists of the two steps:
The first step is the ``salt-bridge formation step'', and
the second step is the ``$\alpha$-helix formation step''.

\end{abstract}

\maketitle

%

  \section{Introduction}
  \label{intro:sec}

In recent years, generalized-ensemble algorithms are frequently employed to study
biomolecular systems (for reviews, see, e.g., Refs.~\cite{mso01,ioo07}).
This is because it is difficult to obtain sufficient sampling in
the conformational space of these systems by
conventional canonical-ensemble simulations
\cite{mrrtt53,hlm82,evans83,nose_mp84,nose_jcp84,hoover85,oio07}.
The canonical simulations tend to get trapped in
a local-minimum free-energy state of the biomolecular systems. 

The replica-exchange method (REM) \cite{hn96,so99} is
one of the most well-known methods among the generalized-ensemble algorithms.
Non-interacting copies (replicas) of a target system are employed in REM.
Different temperatures are assigned to the replicas.
By exchanging the temperatures between the replicas,
random walks of the replicas in the temperature space are realized. 
Accordingly, the simulation can escape from local-minimum states. 
It is easier to perform replica-exchange molecular dynamics (MD) or
Monte Carlo (MC) simulations than to perform multicanonical MD \cite{hoe96,naka97} or
MC simulations \cite{berg91,berg92}, 
although the multicanonical algorithm is also one of
the most well-known generalized-ensemble algorithms.
In the multicanonical and similar algorithms 
\cite{oo04a,oo04b,oo04c,oo04d,oo06,berg03,itoh07a,ok08,ok11,ok12} 
non-Boltzmann weight factors are used as the weight factors. 
These non-Boltzmann weight factors are not {\it a priori} known and 
have to be determined by iterative procedures. 
In REM, however, the usual Boltzmann weight factor is employed for each replica.
Therefore, there is no necessity to perform the procedures of
obtaining the weight factor.

In REM the Metropolis algorithm \cite{mrrtt53} is utilized to
exchange the temperatures between the replicas.
The Metropolis algorithm is a Markov chain Monte Carlo (MCMC) method and
widely employed to obtain a required ensemble.
The Metropolis algorithm is designed so as to
satisfy the detailed balance condition, 
which is a sufficient condition to perform state transitions based on MCMC.
Trials of such state transitions are accepted or rejected stochastically, and
their rejection ratio increases when there is 
a large difference between the probability distributions.
In order to minimize the rejection ratio,
new algorithm was proposed recently by Suwa and Todo \cite{st10}.

In the Suwa-Todo algorithm the detailed balance condition is not imposed.
They introduced a graphical procedure called {\it weight allocation}
instead of solving the detailed balance condition algebraically.
By minimizing the rejection ratio with the weight allocation, this algorithm
realizes efficient sampling of states.
For a system which has only two states for each particle
like the Ising model, however, this algorithm is exactly the same as
the Metropolis algorithm \cite{st10}.

To take advantage of the Suwa-Todo algorithm,
one might consider to exchange temperatures in REM by this algorithm.
For example, let us consider the following exchange of
the temperatures $T_1$ and $T_2$ between the replicas, Replica 1 and Replica 2:
$\left({\rm Replica}~1~{\rm at}~T_1 ; {\rm Replica}~2~{\rm at}~T_2 \right) \rightarrow
\left({\rm Replica}~1~{\rm at}~T_2 ; {\rm Replica}~2~{\rm at}~T_1 \right)$.
In this case, the number of all combinations of
the replicas and temperatures is only two,
$\left({\rm Replica}~1~{\rm at}~T_1 ; {\rm Replica}~2~{\rm at}~T_2 \right)$ and 
$\left({\rm Replica}~1~{\rm at}~T_2 ; {\rm Replica}~2~{\rm at}~T_1 \right)$.
For this exchange, therefore, the Suwa-Todo algorithm is exactly the same as
the Metropolis algorithm.
In general, because temperatures are exchanged 
between two replicas as in this example,
REM is not able to take advantage of the Suwa-Todo algorithm.

In order to solve this difficulty,
we propose a new generalize-ensemble algorithm,
which we refer to as the {\it replica-permutation method} (RPM).
In this method temperatures are permuted among more than two replicas
by the Suwa-Todo algorithm.
We can utilize the Suwa-Todo algorithm
because the number of all combinations of 
the replicas and temperatures is larger than two.

We first apply this new method to particles in a double-well potential energy.
For a comparison purposes,
RPM with the Metropolis algorithm,
and REM are also performed.
As the second application of RPM,
we employ Met-enkephalin in vacuum.
This penta-peptide is often used as a test system 
to see the usefulness of new algorithms
\cite{mho98,itoh04,itoh06,itoh07b}.
Furthermore, RPM is applied to
a C-peptide analog of ribonuclease A in explicit water,
which is known to form a $\alpha$-helix structure
\cite{skbmycsb85,skysb87,obysdw89,fsysb90,lmhjsb98},
to see its sampling efficiency for a larger biomolecular system.
The results of the second and third applications are compared with those of REM.

It is considered that the $\alpha$-helix structure of
the C-peptide analog is stabilized by salt bridges (SBs) between
Gly2 and Arg10 \cite{fsysb90}.
We discuss the role of the SBs for the $\alpha$-helix structure.
Furthermore, a folding pathway based on a free-energy landscape 
is presented from our simulations.

In Section \ref{method:sec} we describe formulation of RPM.
The Suwa-Todo algorithm and the graphical procedure, weight allocation, for
MCMC are also introduced in this section.
We present the details of our simulations in Section \ref{det:sec}. 
The results are shown in Section \ref{rad:sec}.
Section \ref{conc:sec} is devoted to conclusions.

  \section{Methods}
  \label{method:sec}

  \subsection{Markov Chain Monte Carlo Method with the Weight Allocation}
  \label{mc_wa:subsec}

We first describe usual MCMC method with the Metropolis and Suwa-Todo algorithms. 
Let us consider that a system has $n$ states and that
the weight of state $i$ is expressed by $w_i$ ($i=1, \cdots, n$).
In MCMC the weight $w_i$ is given by
\begin{equation}
w_{i} = \sum_{j=1}^{n} w_{j} P(j \rightarrow i)~,
\label{markov_p}
\end{equation}
where $P(j \rightarrow i)$ is the transition probability from state $j$ to state $i$.
By defining the amount of stochastic flow from state $j$ to state $i$ as
\begin{equation}
v(j \rightarrow i) \equiv w_{j} P(j \rightarrow i)~,
\label{def_v}
\end{equation}
Eq.~(\ref{markov_p}) can be rewritten as
\begin{equation}
w_{i} = \sum_{j=1}^{n} v(j \rightarrow i)~.
\label{markov_v}
\end{equation}
This amount of stochastic flow $v(j \rightarrow i)$ also satisfies
\begin{equation}
w_{j} = \sum_{i=1}^{n} v(j \rightarrow i)~,
\label{normalization_v}
\end{equation}
because
\begin{equation}
\sum_{i=1}^{n} P(j \rightarrow i) = 1~.
\label{normalization_p}
\end{equation}
From Eqs.~(\ref{markov_v}) and (\ref{normalization_v}),
the global balance equation is derived:
\begin{equation}
\sum_{i=1}^{n} v(i \rightarrow j) = \sum_{i=1}^{n} v(j \rightarrow i)~.
\label{global_balance}
\end{equation}
By performing state transitions with $v(j \rightarrow i)$
satisfying this equation, a required ensemble is obtained.

In the Metropolis algorithm the detailed balance condition,
which is a sufficient condition for Eq.~(\ref{global_balance}), 
is employed to obtain the required ensemble.
Here, the detailed balance condition is described in terms of
the amount of stochastic flow as follows:
\begin{equation}
v(i \rightarrow j) = v(j \rightarrow i)~.
\label{detailed_balance}
\end{equation}
$v(j \rightarrow i)$ is given by
the following equation so as to satisfy Eq.~(\ref{detailed_balance}):
\begin{equation}
v(j \rightarrow i) = \frac{1}{n-1}
{\rm min}\left[w_{j},w_{i} \right]~, \hspace{5 mm} j \neq i~,
\label{transv_m}
\end{equation}
where the coefficient $1/(n-1)$ comes from a random selection of state $i$
from $(n-1)$ candidates except state $j$.

\begin{figure}
\includegraphics[width=15cm,keepaspectratio]{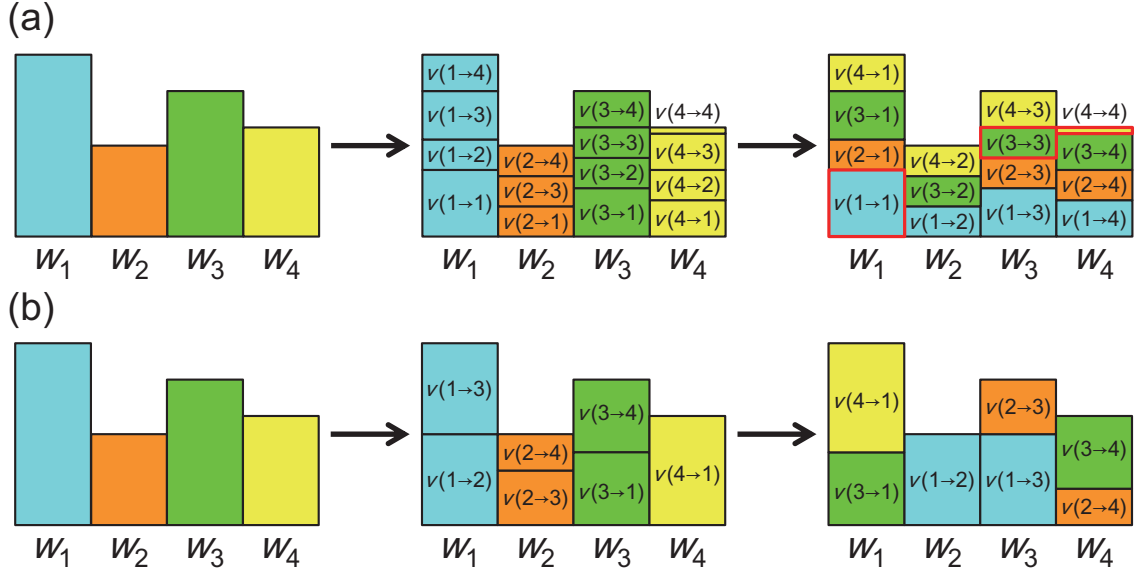}
\caption{
Schematic figures of the weight allocation of the (a) Metropolis and
(b) Suwa-Todo algorithms.
Red frame blocks represent rejected flows 
$v(i \rightarrow i)$ ($i=1,\cdots,4$).
}
\label{box:fig}
\end{figure}

When $v(j \rightarrow i)$ satisfies
Eqs.~(\ref{markov_v}) and (\ref{normalization_v}),
Eq.~(\ref{global_balance}) is automatically fulfilled . 
Therefore, we focus on Eqs.~(\ref{markov_v}) and (\ref{normalization_v}) after this.
These two equations can be understood visually 
by the {\it weight allocation} as follows (see Fig.~\ref{box:fig}(a)):
Equation~(\ref{markov_v}) is regarded as filling a box whose size is $w_i$ by
blocks whose sizes are $v(j \rightarrow i)$ ($j=1, \cdots, n$) 
without any vacant space.
Equation~(\ref{normalization_v}) is regarded as dividing a block 
whose size is $w_j$ into
smaller blocks whose sizes are $v(j \rightarrow i)$ ($i=1, \cdots, n$).
In order to satisfy both equations simultaneously for any $i$ and $j$, therefore,
we prepare blocks $w_j$ and boxes $w_i$ ($i,j=1, \cdots, n$).
By using blocks $v(j \rightarrow i)$ divided from $w_j$,
all $w_i$ are filled without any space.
Figure~\ref{box:fig}(a) shows this weight allocation of Metropolis algorithm
for a system which has four states ($n=4$).
The block size of $v(j \rightarrow i)$ is calculated from Eq.~(\ref{transv_m}).
Red frame blocks represent rejected flow $v(i \rightarrow i)$ ($i=1,\cdots,4$), and
their sizes are associated directly with
the average rejection ratio $\sum_{i}v(i \rightarrow i)/\sum_{i}w_i$.

A new algorithm is proposed recently by Suwa and Todo through
the weight allocation to minimize the average rejection ratio for
state transitions in MCMC.
We refer to this algorithm as the Suwa-Todo algorithm.
In this algorithm $v(j \rightarrow i)$ satisfies
Eqs.~(\ref{markov_v}) and (\ref{normalization_v}) without
imposing the detailed balance condition in Eq.~(\ref{detailed_balance}).
In the Suwa-Todo weight allocation,
$w_j$ is divided and $w_i$ is filled as follows (see Fig.~\ref{box:fig}(b)):
\begin{itemize}
\item[(i)]
Choose the state which has the maximum weight.
If two or more states have the maximum weight,
one of them is chosen.
Here, we assume that $w_1$ is the maximum weight without loss of generality.
\item[(ii)]
Box $w_2$ is filled by block $w_1$ ($v(1 \rightarrow 2)$).
If block $w_1$ still remains after filling box $w_2$,
try to fill the next box $w_3$ ($v(1 \rightarrow 3)$).
this process is continued until the block size of $w_1$ become 0 
($v(1 \rightarrow 4), \cdots, v(1 \rightarrow k)$).
\item[(iii)]
By using block $w_2$, fill boxes in turns from the last partially filled box at Step (ii)
($v(2 \rightarrow k), \cdots, v(2 \rightarrow l)$).
This procedure is repeated for the blocks $w_3, \cdots, w_n$
($v(3 \rightarrow l), \cdots$).
\item[(iv)]
Once all the boxes except $w_1$ are saturated,
box $w_1$ ($\cdots, v(n \rightarrow 1)$) is filled.
\end{itemize}

Figure~\ref{box:fig}(b) shows the weight allocation for $n=4$.
In this figure the average rejection ratio 
$\sum_{i}v(i \rightarrow i)/\sum_{i}w_i$ is 0.
In this Suwa-Todo algorithm,
$v(j \rightarrow i)$ is given by
\begin{equation}
v(j \rightarrow i) = {\rm max} \left[0, {\rm min} \left[
\Delta_{ji}, w_{j}+w_{i}-\Delta_{ji},w_{j},w_{i}
\right] \right]~,
\label{transv_wa}
\end{equation}
where
\begin{equation}
\Delta_{ji} \equiv S_{j}-S_{i-1}+w_{1}~,
\label{delt_ij}
\end{equation}
and
\begin{equation}
S_{i} \equiv \sum_{j=1}^{i} w_{j}~, \hspace{5 mm} S_{0} \equiv S_{n}~.
\label{s_i}
\end{equation}
By performing state transitions based on
the amount of stochastic flow $v(j \rightarrow i)$ or
the transition probability $P(j \rightarrow i) = v(j \rightarrow i)/w_{j}$,
the required ensemble is obtained.

Regarding the rejection ratio, from Eq.~(\ref{transv_wa}),
\begin{equation}
v(i \rightarrow i) = \left\{
\begin{array}{ll}
{\rm max} \left[0, 2w_{1}-S_{n} \right]~, & i=1~, \\
0~, & i \geq 2~.
\end{array}
\right.
\label{reject}
\end{equation}
Therefore, the rejection ratio becomes 0, if
\begin{equation}
w_{1} \leq \frac{S_{n}}{2}~.
\label{w_1}
\end{equation}
It means a reject-free MC simulation is realized 
if $w_{1}$ is less than or equal to the half of 
the total weight $S_{n}=\sum_{j=1}^{n} w_{j}$~.
This condition can be fulfilled in most cases.

  \subsection{Replica-Exchange Method with the Metropolis Algorithm}
  \label{rem:subsec}

We consider a system of {\it N} atoms with their coordinate vectors and 
momentum vectors denoted by 
$q \equiv \left\{ \mbox{\boldmath $q$}_{1},\cdots,\mbox{\boldmath $q$}_{N} \right\}$ and 
$p \equiv \left\{ \mbox{\boldmath $p$}_{1},\cdots,\mbox{\boldmath $p$}_{N} \right\}$,
respectively. 
The Hamiltonian $H$ in state $x \equiv (q,p)$ is given by
the sum of the kinetic energy {\it K} and potential energy $V$: 
\begin{eqnarray} 
H(x) = K(p)+V(q)~.
\label{hamiltonian}
\end{eqnarray}
In the canonical ensemble at temperature $T$,  
each state $x$ is weighted by the Boltzmann factor:
\begin{equation}
W_{\rm B}(x) = e^{-\beta H(x)}~,
\label{wcano}
\end{equation}
where $\beta = 1/k_{\rm B} T$ ($k_{\rm B}$ is the Boltzmann constant). 

Let us suppose that there are $M$ {\it non-interacting} copies (or replicas) of
the original system in the canonical ensemble at
$M$ different temperatures $T_m$ ($m=1, \cdots, M$).
In REM the replicas are arranged so that there would be always
exactly one replica at each temperature.
In other words, there is a one-to-one correspondence between
the replicas and temperatures.
Therefore, the label $i$ ($i=1, \cdots, M$) for the replicas is
a permutation of the label $m$ ($m=1, \cdots, M$) for the temperatures, and vice versa:
\begin{equation}
\left\{
\begin{array}{rl}
i &=~ i(m) ~\equiv~ f(m)~, \\
m &=~ m(i) ~\equiv~ f^{-1}(i)~,
\end{array}
\right.
\label{permu}
\end{equation}
where $f(m)$ is a permutation function of $m$ and $f^{-1}(i)$ is its inverse.

Let $X_{\alpha} = \left\{x_1^{[i(1)]}, \cdots, x_M^{[i(M)]} \right \} 
= \left\{x_{m(1)}^{[1]}, \cdots, x_{m(M)}^{[M]} \right \}$ 
stands for a ``state'' in REM.
Here, the superscript $i$ and the subscript $m$ in $x_m^{[i]}$ are labels of
the replicas and temperatures, respectively.
All possible combinations between the replicas and temperatures are labeled by
the subscript $\alpha$.
The state $X_{\alpha}$ is specified by the $M$ sets of coordinates $q^{[i]}$ and
momenta $p^{[i]}$ of $N$ atoms in replica $i$ at temperature $T_m$:
\begin{equation}
x_m^{[i]} \equiv \left(q^{[i]},p^{[i]}\right)_m~.
\label{state1}
\end{equation}
Because the replicas are non-interacting, the weight factor $w_{\rm R}(X_{\alpha})$ for
the state $X_{\alpha}$ is given by
the product of Boltzmann factors for each replica $i$ (or at temperature $T_m$):
\begin{equation}
w_{\rm R} \left(X_{\alpha} \right) = \displaystyle{\prod_{i=1}^M}
\displaystyle{\exp \left\{ 
- \beta_{m(i)} H\left( x_{m(i)}^{[i]} \right) \right\} }
 = \displaystyle{\prod_{m=1}^M} \displaystyle{\exp \left\{
- \beta_{m} H\left( x_m^{[i(m)]} \right) \right\}~, }
\label{whrem}
\end{equation}
where $i(m)$ and $m(i)$ are the permutation functions in Eq.~(\ref{permu}).

We now consider exchanging a pair of replicas in REM.
Suppose we exchange replicas $j$ and $k$
which are at temperatures $T_m$ and $T_n$, respectively ($j=i(m)$, $k=i(n)$):  
\begin{equation}
X_{\alpha} = \left\{\cdots, x_m^{[j]}, \cdots, x_n^{[k]}, \cdots \right\} 
\longrightarrow \ 
X_{\beta} = \left\{\cdots, x_m^{[k]}, \cdots, x_n^{[j]}, 
\cdots \right\}~. 
\label{rep_exchange}
\end{equation}
From Eq.~(\ref{transv_m})
the amount of stochastic flow $v\left(X_{\alpha} \rightarrow X_{\beta} \right)$ 
for this replica exchange is given by
\begin{equation}
v \left(X_{\alpha} \rightarrow X_{\beta} \right) =
C{\rm min}\left[w_{\rm R} \left( X_{\alpha} \right),
w_{\rm R} \left( X_{\beta} \right) \right]~,
\label{transv_m_rem}
\end{equation}
and the transition probability $P\left(X_{\alpha} \rightarrow X_{\beta} \right)$ is
expressed by
\begin{equation}
P\left(X_{\alpha} \rightarrow X_{\beta} \right) =
C{\rm min}\left[1,
\frac{w_{\rm R} \left( X_{\beta} \right)}
{w_{\rm R} \left( X_{\alpha} \right)} \right]~,
\label{transp_m_rem}
\end{equation}
where $C = 1/\left({}_{M} C_{2} \right)$ if
replicas $j$ and $k$ were selected randomly among $M$ replicas.
Here, ${}_{M} C_{2}$ is the number of 2-combinations from $M$ elements:
${}_{M} C_{2} = M!/\{ (M-2)! 2! \}$~.
In the usual REM, replica exchanges are tried only between neighboring 
two temperatures.
In this case, $C=1$ because $M=2$.

  \subsection{Replica-Permutation Method with the Suwa-Todo Algorithm}
  \label{rpm:subsec}

We consider to perform a replica permutation among all $M$ replicas
as a generalization of a replica exchange in Eq.~(\ref{rep_exchange}):
\begin{equation}
X_{\alpha} = \left\{x_1^{[i(1)]}, \cdots, x_M^{[i(M)]} \right\} 
\longrightarrow \ 
X_{\beta} = \left\{ x_1^{[j(1)]}, \cdots, x_M^{[j(M)]} \right\}~,
\label{rep_permutation}
\end{equation}
where both $i$ and $j$ are permutation functions and $i \neq j$.
Not only an exchange between two replicas, 
but also a permutation among more than two replicas are 
allowed in this method. 
Note that the number of all possible combinations between
the replicas and temperatures is $M!$.
Therefore, the index $\alpha$ of $X_{\alpha}$ takes a value between 1 and $M!$.

In the Metropolis algorithm,
the transition probability $P\left(X_{\alpha} \rightarrow X_{\beta} \right)$ for
this replica permutation is also given by Eq.~(\ref{transp_m_rem}).
The replicas are allowed to transit to non-neighboring temperatures,
but $P\left(X_{\alpha} \rightarrow X_{\beta} \right)$ takes a quite small value for
such replica permutation.
Accordingly, most of such replica permutations are rejected.
The number of replica permutations in which any replica does not transit to
non-neighboring temperatures is given by
\begin{equation}
\sum_{n=1}^{\left[ \frac{M}{2} \right]} {}_{M-n} C_{n}~,
\label{reject_permulation}
\end{equation}
where 
\begin{equation}
\left[ \frac{M}{2} \right]
= \left\{
\begin{array}{ll}
\frac{M  }{2}~, & {\rm for~even~number~of}~M~, \\
\frac{M-1}{2}~, & {\rm for~odd~number~of}~M~.
\end{array}
\right.
\label{M2:eq}
\end{equation}
On the other hand, 
the number of all replica-permutation candidates is $M!-1$
because the current combination between the replicas and temperatures is not included.
Therefore, the probability of trying replica permutations which
do not include non-neighboring transitions is
$\left( \sum_{n=1}^{\left[ \frac{M}{2} \right]} {}_{M-n} C_{n} \right)/(M!-1)$.
Thus, most of the replica-permutation trials are rejected
for large $M$, because
\begin{equation}
\sum_{n=1}^{\left[ \frac{M}{2} \right]} {}_{M-n} C_{n} <
\sum_{n=0}^{M} {}_{M} C_{n} = 2^M \ll M!~.
\label{reject_prob}
\end{equation}

To avoid this rejection problem,
we apply the Suwa-Todo algorithm to the replica permutations.
As in Sec.~\ref{mc_wa:subsec},
we assume that $w_{\rm R} \left( X_{1} \right)$ is the maximum weight 
without loss of generality.
The amount of stochastic flow $v\left(X_{\alpha} \rightarrow X_{\beta} \right)$ is
determined by the weight allocation in the same way also as in Sec.~\ref{mc_wa:subsec}
only by replacing the weight $w_{i}$ to $w_{\rm R} \left( X_{\alpha} \right)$.
From Eqs.~(\ref{transv_wa}), (\ref{delt_ij}), and (\ref{s_i}),
$v\left(X_{\alpha} \rightarrow X_{\beta} \right)$ is given by
\begin{equation}
v\left(X_{\alpha} \rightarrow X_{\beta}\right) = {\rm max} \left[0, {\rm min} \left[
\Delta_{\alpha \beta}, w_{\rm R}\left(X_{\alpha} \right)+
w_{\rm R}\left(X_{\beta} \right)-\Delta_{\alpha \beta},
w_{\rm R}\left(X_{\alpha} \right),w_{\rm R}\left(X_{\beta}\right)
\right] \right]~,
\label{transv_wa_rpm}
\end{equation}
where
\begin{equation}
\Delta_{\alpha \beta} \equiv S_{\alpha}-S_{\beta-1}+
w_{\rm R}\left(X_{1} \right)~,
\label{delt_ab}
\end{equation}
and
\begin{equation}
S_{\alpha} \equiv \sum_{\beta=1}^{\alpha} w_{\rm R}\left(X_{\beta} \right)~,
\hspace{5 mm} S_{0} \equiv S_{M!}~.
\label{s_a}
\end{equation}
If $w_{\rm R} \left( X_{\gamma} \right)$ ($\gamma \neq 1$) 
is the maximum weight more generally,
Eqs.~(\ref{delt_ab}) and (\ref{s_a}) are modified as follows:
\begin{equation}
\Delta_{\alpha \beta} \equiv S_{\alpha}-S_{\beta-1}+
w_{\rm R}\left(X_{\gamma} \right)~,
\label{delt_ab_gene}
\end{equation}
and
\begin{eqnarray}
S_{\alpha} &\equiv& \left\{
\begin{array}{ll}
\displaystyle
\sum_{\beta=\gamma}^{\alpha} w_{\rm R}\left(X_{\beta} \right)~,
{\rm for~} \alpha \ge \gamma~, \\
\displaystyle
\sum_{\beta=\gamma}^{M!} w_{\rm R}\left(X_{\beta} \right)+
\sum_{\beta=1}^{\alpha} w_{\rm R}\left(X_{\beta} \right)~,
{\rm for~} \alpha < \gamma~, \\
\end{array}
\right. \\
S_{0} &\equiv& S_{M!}~. \nonumber
\label{s_a_gene}
\end{eqnarray}

A replica-permutation simulation with
the Suwa-Todo algorithm is performed as follows:
\begin{itemize}
\item[Step 1:]
The label $\alpha$ ($\alpha=1,\cdots,M!$) of $X_{\alpha}$ is assigned to
all combinations between the replicas and temperatures.
Table~\ref{ex_state:table} shows an example for $M=3$ (three replicas).
\item[Step 2:]
For each replica, a canonical MD or MC simulation at
the assigned temperature is carried out simultaneously and
independently for a certain steps. 
\item[Step 3:]
A replica-permutation trial is performed as follows:
First, each weight is obtained by Eq.~(\ref{whrem}),
and the maximum weight $w_{\rm R} \left( X_{\gamma} \right)$ is determined. 
Next, we calculate the amount of stochastic flow
$v\left(X_{\alpha} \rightarrow X_{\beta} \right)$ in Eq.~(\ref{transv_wa_rpm}) and
the transition probability 
$P \left(X_{\alpha} \rightarrow X_{\beta} \right) = 
v \left(X_{\alpha} \rightarrow X_{\beta} \right)/w_{\rm R} \left( X_{\alpha} \right)$ for
$\beta=1,\cdots,M!$.
Finally, transition from State $X_{\alpha}$ to
State $X_{\beta}$ is accepted stochastically with the probability 
$P\left(X_{\alpha} \rightarrow X_{\beta} \right)$.
\end{itemize}
Repeating Step 2 and Step 3, we can carry out the replica permutation 
MD or MC simulation.

\begin{figure}
\includegraphics[width=15cm,keepaspectratio]{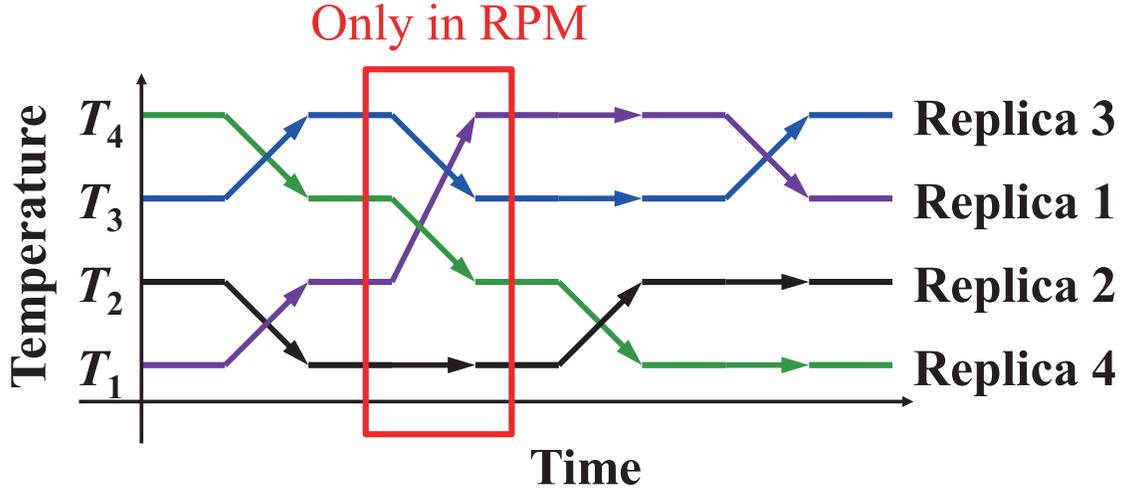}
\caption{
An example of time series of temperatures in RPM.
The transitions of replicas in the red square frame is not realized in REM.
}
\label{time_rep:fig}
\end{figure}

Figure~\ref{time_rep:fig} shows an example of time series of temperatures in RPM.
This method realizes not only minimization of the rejection ratio
but also transitions of the replicas to non-neighboring temperatures. 

The number of combinations between
the replicas and temperatures increases in proportion to $M!$.
For a large number of replicas, 
we can divide all replicas and temperatures into subsets
to decrease the number of combinations. 
Although such a division is not necessary, 
three to eight replicas are appropriate in each subset.
As an example, let us consider that the total number of the replicas is eight and
that they are divided into two subsets which have four replicas.
In this case, the replica-permutation simulation is performed as follows:
\begin{itemize}
\item[Step 1:]
Let us suppose that the temperatures are assigned to the replicas 
at an initial state as
\begin{eqnarray}
\left(
\begin{array}{c}
{\rm Replica}~1~{\rm at}~T_1 \\
{\rm Replica}~2~{\rm at}~T_2 \\
\vdots                       \\
{\rm Replica}~8~{\rm at}~T_8 \\
\end{array}
\right)~.
\nonumber
\end{eqnarray}
They are divided into two subsets:
\begin{eqnarray}
\left(
\begin{array}{c}
{\rm Replica}~1~{\rm at}~T_1 \\
{\rm Replica}~2~{\rm at}~T_2 \\
{\rm Replica}~3~{\rm at}~T_3 \\
{\rm Replica}~4~{\rm at}~T_4 \\
\end{array}
\right)~{\rm and}~\left(
\begin{array}{c}
{\rm Replica}~5~{\rm at}~T_5 \\
{\rm Replica}~6~{\rm at}~T_6 \\
{\rm Replica}~7~{\rm at}~T_7 \\
{\rm Replica}~8~{\rm at}~T_8 \\
\end{array}
\right)~.
\nonumber
\end{eqnarray}
All combinations between the replicas and temperatures for the former subset
and those for the latter subset are labeled by $X^1_{\alpha}$ and
$X^3_{\alpha}$ ($\alpha=1,\cdots,4!$), respectively.
Moreover, two more subsets are prepared so that components would differ from
those of the previous subsets:
\begin{eqnarray}
\left(
\begin{array}{c}
{\rm Replica}~3~{\rm at}~T_3 \\
{\rm Replica}~4~{\rm at}~T_4 \\
{\rm Replica}~5~{\rm at}~T_5 \\
{\rm Replica}~6~{\rm at}~T_6 \\
\end{array}
\right)~{\rm and}~\left(
\begin{array}{c}
{\rm Replica}~7~{\rm at}~T_7 \\
{\rm Replica}~8~{\rm at}~T_8 \\
{\rm Replica}~1~{\rm at}~T_1 \\
{\rm Replica}~2~{\rm at}~T_2 \\
\end{array}
\right)~.
\nonumber
\end{eqnarray}
The labels for the former subset and the latter subset are
$X^2_{\alpha}$ and $X^4_{\alpha}$, respectively.
\item[Step 2:]
For each replica, a canonical MD or MC simulation at
the assigned temperature is carried out simultaneously and
independently for a certain steps. 
\item[Step 3:]
At an odd number of trail time,
replica permutations for $X^1_{\alpha}$ and $X^3_{\alpha}$ are carried out.
That is, 
four replicas are permutated among four corresponding temperatures
in each subset. 
At an even number of trial time,
replica permutations for $X^2_{\alpha}$ and $X^4_{\alpha}$ are performed.
\item[Step 4:]
As a result of the replica permutation for $X^1_{\alpha}$, 
let us suppose that the combination of the replicas 
and temperatures is changed as
\begin{equation}
\left(
\begin{array}{c}
{\rm Replica}~1~{\rm at}~T_1 \\
{\rm Replica}~2~{\rm at}~T_2 \\
{\rm Replica}~3~{\rm at}~T_3 \\
{\rm Replica}~4~{\rm at}~T_4 \\
\end{array}
\right)
\rightarrow
\left(
\begin{array}{c}
{\rm Replica}~1~{\rm at}~T_1 \\
{\rm Replica}~3~{\rm at}~T_2 \\
{\rm Replica}~2~{\rm at}~T_3 \\
{\rm Replica}~4~{\rm at}~T_4 \\
\end{array}
\right)~.
\label{rpm_ex1}
\end{equation}
Due to this permutation,
the combination in $X^2_{\alpha}$ is automatically changed without
a replica permutation for this subset:
\begin{equation}
\left(
\begin{array}{c}
{\rm Replica}~3~{\rm at}~T_3 \\
{\rm Replica}~4~{\rm at}~T_4 \\
{\rm Replica}~5~{\rm at}~T_5 \\
{\rm Replica}~6~{\rm at}~T_6 \\
\end{array}
\right)
\stackrel{\mbox{by Eq.~(\ref{rpm_ex1})}}{\hbox to 25mm{\rightarrowfill}}
\left(
\begin{array}{c}
{\rm Replica}~2~{\rm at}~T_3 \\
{\rm Replica}~4~{\rm at}~T_4 \\
{\rm Replica}~5~{\rm at}~T_5 \\
{\rm Replica}~6~{\rm at}~T_6 \\
\end{array}
\right)~.
\label{rpm_ex2}
\end{equation}
To avoid re-labeling for $X^2_{\alpha}$, 
we perform the next permutation 
regarding Replica~2 as Replica~3.
Although the combination of the temperatures and replicas may be changed by
the last replica permutation, we do not need to relabel the combinations as
in Table~\ref{ex_state:table} by replacing the replicas in this way.
\end{itemize}
Repeating Step 2 to Step 4, 
we continue the replica permutation MD or MC simulation.

  \subsection{Reweighting Techniques}
  \label{reweight:subsec}

The results obtained from RPM can be analyzed by
the reweighting techniques as in REM \cite{fs89,kbskr92}. 
Let us suppose that we have carried out a RPM simulation with
$M$ replicas and $M$ different temperatures $T_m$ ($m=1,\cdots,M$).

For appropriate reaction coordinates $\xi_{1}$ and $\xi_{2}$, 
the canonical probability distribution $P_{T}(\xi_{1},\xi_{2})$ at
any temperature $T$ can be calculated from
\begin{equation}
P_{T}(\xi_{1},\xi_{2}) = \sum_{E}
\frac{\displaystyle \sum^{M}_{m=1} \left( g_{m} \right)^{-1} 
N_{m}(E;\xi_{1},\xi_{2}) e^{-\beta E}}
{\displaystyle \sum^{M}_{m=1} \left( g_{m} \right)^{-1} 
n_{m} e^{f_{m}-\beta_{m} E}}~,
\label{prob}
\end{equation}
and 
\begin{equation}
e^{-f_{m}} = \sum_{\xi_{1},\xi_{2}} P_{T_m}(\xi_{1},\xi_{2})~.
\label{dless_free}
\end{equation}
Here, $g_{m}=1+2 \tau_{m}$, $\tau_{m}$ is the integrated autocorrelation time at
temperature $T_m$, $N_{m}(E;\xi_{1},\xi_{2})$ is the histogram of the potential energy and
the reaction coordinates $\xi_{1}$ and $\xi_{2}$ at $T_m$,
and $n_{m}$ is the total number of samples obtained at $T_m$.
Note that this probability distribution is not normalized.
Equations (\ref{prob}) and (\ref{dless_free}) are solved self-consistently by iteration. 
For biomolecular systems the quantity $g_m$ can safely be set
to be a constant in the reweighting formulas \cite{kbskr92},
and so we set $g_m =1$ throughout the analyses in the present work.
These equations can be easily generalized 
to any reaction coordinates $(\xi_{1},\xi_{2},\cdots)$. 

From the canonical probability distribution $P_{T}(\xi_{1},\xi_{2})$ in
Eq.~(\ref{prob}), the expectation value of a physical quantity $A$ at
any temperature $T$ is given by 
\begin{equation}
\left<A \right>_{T} = 
\frac{\displaystyle \sum_{\xi_{1},\xi_{2}} A(\xi_{1},\xi_{2})
P_{T}(\xi_{1},\xi_{2})}
{\displaystyle \sum_{\xi_{1},\xi_{2}} P_{T}(\xi_{1},\xi_{2})}~.
\label{rew_ev}
\end{equation}
We can also calculate the free energy (or, the potential of mean force) as
a function of the reaction coordinates $\xi_{1}$ and $\xi_{2}$ at any temperature $T$ from
\begin{equation}
F_{T}(\xi_{1},\xi_{2}) = -k_{\rm B}T{\rm ln}{P_{T}(\xi_{1},\xi_{2})}~.
\label{def_free}
\end{equation}

  \section{Computational Details}
  \label{det:sec}

  \subsection{Asymmetric Double-Well Potential Energy}
  \label{dwell_com:subsec}

In order to verify that RPM gives correct ensembles,
we first applied this method to a simple system.
The system has 100 non-interacting particles
with a one-dimensional asymmetric double-well potential energy.
The potential energy $V\left(q \right)$ at a coordinate $q$ is defined by
\begin{equation} 
V\left(q \right) = \left(\left(q+1 \right)^{2} -1 \right)
\left(\left(q-1 \right)^{2} -0.9 \right)~
{\rm kcal}/({\rm mol} \cdot \mbox{\AA$^4$})~.
\label{double_well}
\end{equation}
To see usefulness of the Suwa-Todo algorithm for the replica-permutation method,
we performed replica-permutation MD simulations with
both Metropolis and Suwa-Todo algorithms.
From now on, we will call the replica-permutation method with the Suwa-Todo algorithm
RPM and that with the Metropolis algorithm M-RPM.
The MD versions of the former and latter are called RPMD and M-RPMD, respectively.
Conventional replica-exchange MD (REMD) simulations were also carried out for
a comparison purposes.
We prepared 40 different initial conditions (ICs) for each method.
The MD simulation from each IC was performed for 10.0 ns 
including equilibration run for 1.0 ns. 
The time step was taken to be 1.0 fs.
The mass of each particle was 1.0 a.u.
Six replicas were used, and
temperatures were set at 200 K, 235 K, 275 K, 325 K, 380 K, and 450 K.
The temperatures were controlled by 
the Gaussian constraint method \cite{hlm82,evans83} to avoid 
the problem of non-ergodicity in 
the Nos\'e-Hoover thermostat \cite{nose_mp84,nose_jcp84,hoover85} 
for the non-interacting particle system.
The trajectory data were stored every 10.0 fs. 
Replica permutations or exchanges were tried every 1.0 ps.

  \subsection{Met-Enkephalin in Vacuum}
  \label{enke_com:subsec}

To see the usefulness of RPM for biomolecular systems, 
we next employed a Met-enkephalin molecule in vacuum as a test system.
The results were compared with those obtained from a REMD simulation.
The AMBER parm99SB force field \cite{amber,amber99sb} was used.
The N-terminus and the C-terminus of Met-enkephalin were blocked by 
the acetyl group and the N-methyl group, respectively.
Therefore, the amino-acid sequence is Ace-YGGFM-Nme.
The SHAKE algorithm \cite{shake} was employed to
constrain bond lengths with the hydrogen atoms during our simulations. 
The temperature was controlled by the Nos\'e-Hoover thermostat
\cite{nose_mp84,nose_jcp84,hoover85}.
The time step was taken to be 1.0 fs. 
The initial conformation was an extended structure.

The number of replicas was 12, and
temperatures were 200 K, 230 K, 265 K, 300 K, 340 K, 385 K,
435 K, 490 K, 555 K, 635 K, 720 K, and 820 K.
For PRM, we divided the 12 replicas into two subsets,
which had six replicas and six temperatures. 
The RPMD and REMD simulations were performed for 50.0 ns per replica
including equilibration run for 1.0 ns. 
The trajectory data were stored every 100 fs. 
Replica permutations or exchanges were tried every 1.0 ps.

  \subsection{C-peptide in Explicit Water}
  \label{cpep_com:subsec}

In order to demonstrate sampling efficiency of RPM for a lager system,
we performed a RPMD simulation of a C-peptide analog 
in explicit water solvent~\cite{skysb87}.
This peptide is known to form an $\alpha$-helix structure 
at a lower temperature than 318 K at pH 5.2 \cite{skysb87,lmhjsb98}.
Because the charges at the peptide termini 
affect helix stability \cite{skbmycsb85},
we blocked the termini of C-peptide by neutral Nme and Ace.
Therefore, the amino-acid sequence is Ace-AETAAAKFLRAHA-Nme.
The histidine residue was protonated to conform our
simulation conditions to the experimental pH.
A REMD simulation was also carried out.
The number of water molecules was 1800, and
two chlorine ions were added as counter ions.
The AMBER parm99SB \cite{amber,amber99sb} was used.
The model for the water molecules was the TIP3P rigid-body model \cite{tip3p}.
Temperature was controlled by 
the Nos\'e-Hoover thermostat \cite{nose_mp84,nose_jcp84,hoover85}.
The SHAKE algorithm \cite{shake} was employed to
constrain bond lengths with the hydrogen atoms of C-peptide and
to fix the water molecule structures during our simulations. 
This system was put in a cubic unit cell with the side length of 38.6 \AA~with
the periodic boundary conditions.
The cutoff distance for the Lennard-Jones potential energy was 12.0 \AA.
The electrostatic potential energy was calculated by the Ewald method \cite{ewald}.
The multiple-time-step method \cite{mts} was employed in our MD simulations.
For interactions between the C-peptide atoms and those between
the C-peptide atoms and the water atoms, the time step was taken to be 1.0 fs.
The time step of 4.0~fs was used for interactions between the water atoms.
The initial conformation was an extended structure.

The numbers of replicas in the RPMD and REMD simulations were 24, and
temperatures were
281 K, 285 K, 289 K, 294 K, 299 K, 304 K, 309 K, 314 K, 320 K, 326 K, 332 K, 338 K,
344 K, 351 K, 358 K, 365 K, 372 K, 380 K, 388 K, 396 K, 405 K, 414 K, 423 K, and 433 K.
These replicas were divided into four subsets in RPM.
Each subset had six replicas and six temperatures. 
The RPMD and REMD simulations were performed for 40.0 ns per replica
including equilibration run for 4.0 ns.
Namely, the production run of each simulation was carried out for 864.0 ns in total.
The trajectory data were stored every 400 fs. 
Trials of replica permutations and exchanges were performed every 4.0 ps.

  \section{Results and Discussion}
  \label{rad:sec}

  \subsection{Asymmetric Double-Well Potential Energy}
  \label{dwell_rd:subsec}

\begin{figure}
\includegraphics[width=13cm,keepaspectratio]{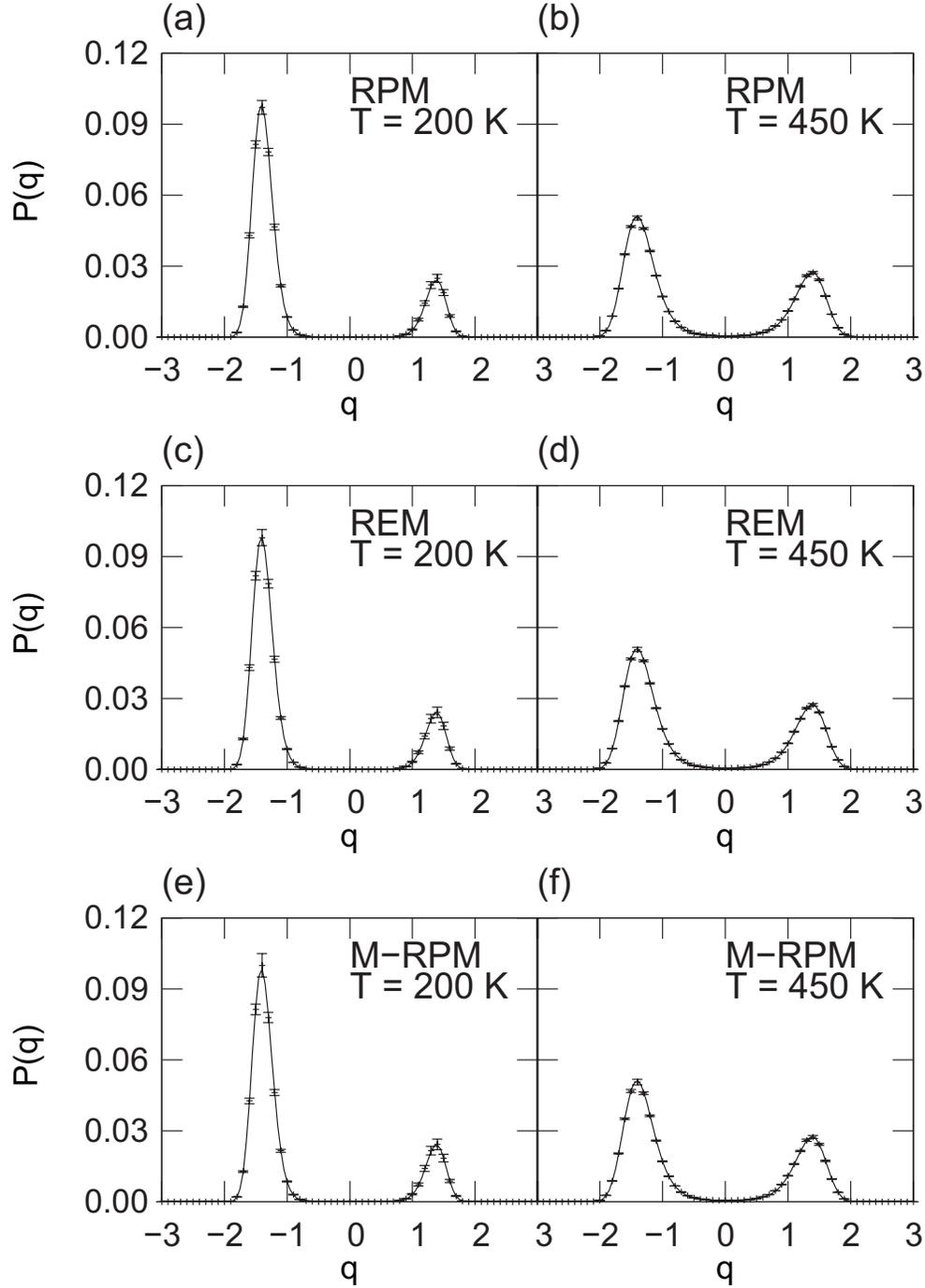}
\caption{
The probability distributions $P\left(q \right)$ of the coordinates $q$ 
at (a), (c), (e) $T=200~{\rm K}$ and (b), (d), (f) $T=450~{\rm K}$. 
These results were obtained from the 
(a), (b) RPMD simulations, 
(c), (d) REMD simulations, and 
(e), (f) M-RPMD simulations.
The solid lines are the probability distributions in Eq.~(\ref{prob_q}).
}
\label{rawhist_dw:fig}
\end{figure}

We first show that RPM (and M-RPM) gives correct probability distributions for
the asymmetric double-well potential energy.
The probability distributions $P\left(q \right)$ in
the RPMD, M-RPMD, and REMD simulations are presented in Fig.~\ref{rawhist_dw:fig}.
Here, $P\left(q \right)$ for each method was obtained from
the 40 simulations starting from the different ICs.
The bin size $\mit\Delta q$ for $P\left(q \right)$ was taken to be 0.05 \AA.
The errors were estimated as the standard deviations of the 40 simulations. 
To see the accuracy of the simulation results,
exact probability distributions $P_{\rm exact}\left(q \right)$ are 
also illustrated as the solid lines in the figure.
$P_{\rm exact}\left(q \right)$ at a temperature $T_{0}$ was calculated numerically by
\begin{equation}
P_{\rm exact}\left(q \right) = C_{\rm DW}
\int_{q-\frac{\mit\Delta q}{2}}^{q+\frac{\mit\Delta q}{2}} dq^{\prime}
{\rm exp}\left( - \beta_{0} V \left(q^{\prime} \right) \right)~,
\label{prob_q}
\end{equation}
where $C_{\rm DW}$ is the normalization constant: 
$C_{\rm DW}= \left( \int_{-\infty}^{\infty} dq P_{\rm exact}\left(q \right) \right)^{-1}$~.
The integration in this equation was computed by the Simpson's method.
As shown in Fig.~\ref{rawhist_dw:fig}, 
all $P\left(q \right)$ show good agreement with
$P_{\rm exact}\left(q \right)$ at all temperatures.
Correct probability distributions are obtained by
not only REM but also RPM (and M-RPM).

\begin{figure}
\includegraphics[width=8cm,keepaspectratio]{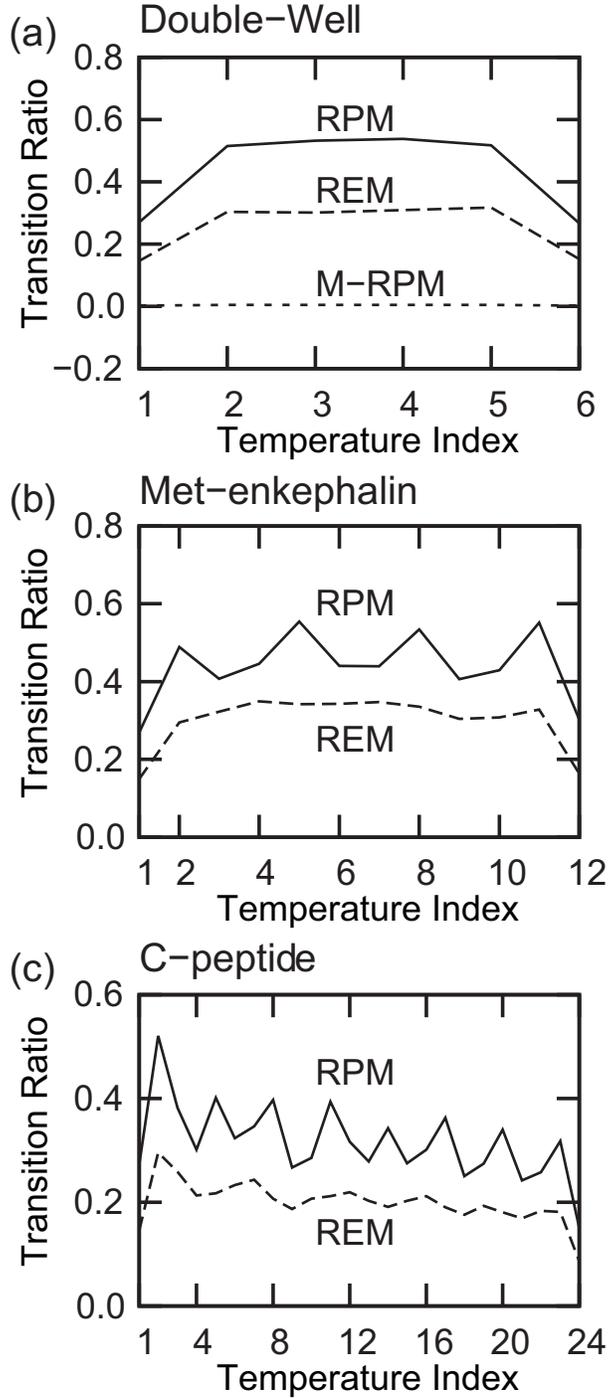}
\caption{
Transition ratios of the replicas for 
(a) particles in the double-well potential, 
(b) Met-enkephalin in vacuum, and 
(c) C-peptide in explicit water. 
Temperatures are represented as the temperature indices.
The smallest and the highest indices correspond
the lowest and the highest temperatures, respectively.
}
\label{trans_ratio:fig}
\end{figure}

The transition ratios of the replicas from a temperature to another temperature in
each method are shown in Fig.~\ref{trans_ratio:fig}(a).
The transition ratio of the replicas is defined here as 
a probability with which a replica at the temperature is 
transferred to another temperature. 
RPM has the largest transition ratios at all temperatures
among all methods.
On the other hand, those of M-RPM are extremely small (the values are 0.003--0.004).
This is because most of the replica-permutation trials were rejected.
To realize efficient replica permutations, therefore,
it is essential to employ the Suwa-Todo algorithm.
The total numbers of tunneling times of all replicas during the simulations
are listed in Table~\ref{tunnel:table}.
Here, when a replica makes a round trip between the lowest and highest temperatures,
it is counted as one tunneling.
The number of the tunneling times is a useful information to see
how efficiently the simulation samples the temperature space.
The value for each method in the table was obtained by taking an average of
the 40 simulations' results.
The number of the tunneling times in RPM was 1.7 times larger than that in REM.
It means that RPM realizes efficient sampling in the temperature space.
On the other hand, M-RPM did not have such a large number of the tunneling times.
This is because its transition ratios were very small.

\begin{figure}
\includegraphics[width=13cm,keepaspectratio]{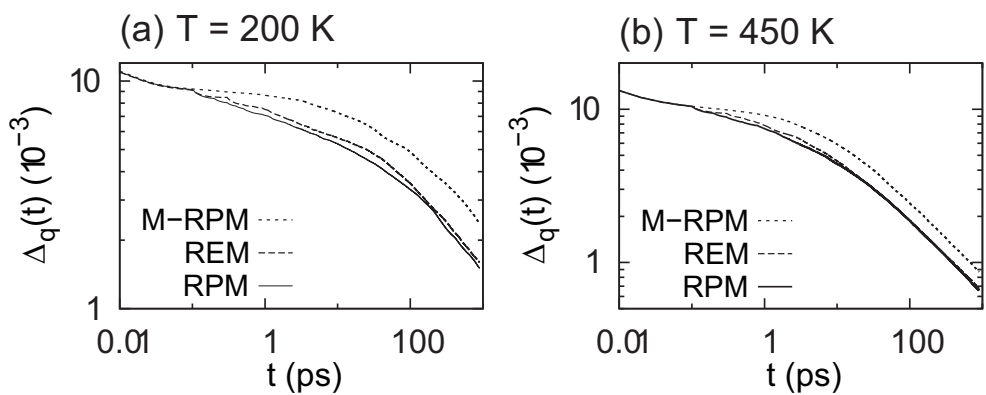}
\caption{
Average deviation $\Delta_q (t)$ in Eq.~(\ref{conv_dw}) 
at (a) $T=200~{\rm K}$ and (b) $T=450~{\rm K}$.
The solid line, the dashed line, and the dotted line show $\Delta_q (t)$ obtained from
the RPMD simulations, the REMD simulations, and the M-RPMD simulations, respectively.
}
\label{time_converge_particle:fig}
\end{figure}

To estimate the convergence speed to the exact probability distributions,
we examined the time series of the average deviations of 
probability distributions $P^{i,j}_{\rm particle}\left(q;t \right)$ 
from $P_{\rm exact}\left(q \right)$,
where $P^{i,j}_{\rm particle}\left(q;t \right)$ is a probability distribution 
obtained from a single MD simulation of particle $i$ from IC $j$ 
accumulated until time $t$. 
The average deviation $\Delta_{q} (t)$ at the time $t$ is defined by
\begin{equation}
\Delta_{q} (t) = \frac{1}{N_{\rm bin}N_{\rm IC}N_{\rm particle}}
\sum^{N_{\rm bin}}_{k=1} \sum^{N_{\rm IC}}_{j=1} \sum^{N_{\rm particle}}_{i=1}
\left| P^{i,j}_{\rm particle}\left(q_k;t \right) -
P_{\rm exact}\left(q_k \right) \right|~,
\label{conv_dw}
\end{equation}
where $N_{\rm bin}$ is the number of the bins,
$N_{\rm IC}$ is the number of ICs,
$N_{\rm particle}$ is the number of the particles in a single simulation,
and $q_k$ is the coordinate at bin $k$.
120 bins ($N_{\rm bin}=120$) ranging between 
$q_k=-3$ and $q_k=3$ were taken into account. 
Figure~\ref{time_converge_particle:fig} shows $\Delta_{q} (t)$ 
calculated from each method.
Convergence of M-RPM is the slowest due to
the low transition ratios of the replicas.
RPM shows slightly faster convergence than REM 
at the lowest temperature $(T=200~{\rm K})$ 
although convergence at highest temperature $(T=450~{\rm K})$ is almost the same.
It is important to increase the sampling efficiency 
at a low temperature because the conformational sampling at a high temperature 
is originally easier than at a low temperature. 
Therefore, RPM realizes efficient sampling not only in the temperature space
but also in the coordinate space at a low temperature. 
It is again because the transition ratios of the replicas by RPM are higher 
than those by REM.

  \subsection{Met-Enkephalin in Vacuum}
  \label{enke_rd:subsec}

The replica-transition ratios and the total numbers of tunneling times
for Met-enkephalin are shown in
Fig.~\ref{trans_ratio:fig}(b) and Table~\ref{tunnel:table}, respectively.
The transition ratios in RPM are larger than those of REM at all temperatures.
RPM has also larger tunneling times than REM.
Thus, efficient sampling in the temperature space was realized by RPM 
for this biomolecular system, too. 

\begin{figure}
\includegraphics[width=10cm,keepaspectratio]{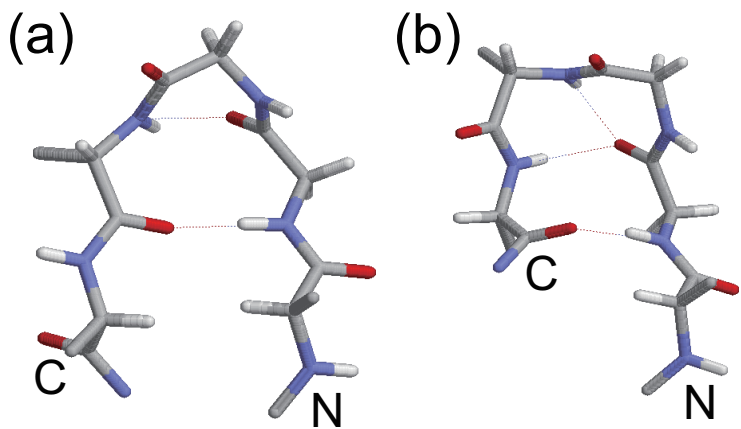}
\caption{
The (a) global-minimum and (b) local-minimum structures of Met-enkephalin in vacuum
for the CHARMM22 force field.
}
\label{reference_enke:fig}
\end{figure}

\begin{figure}
\includegraphics[width=9cm,keepaspectratio]{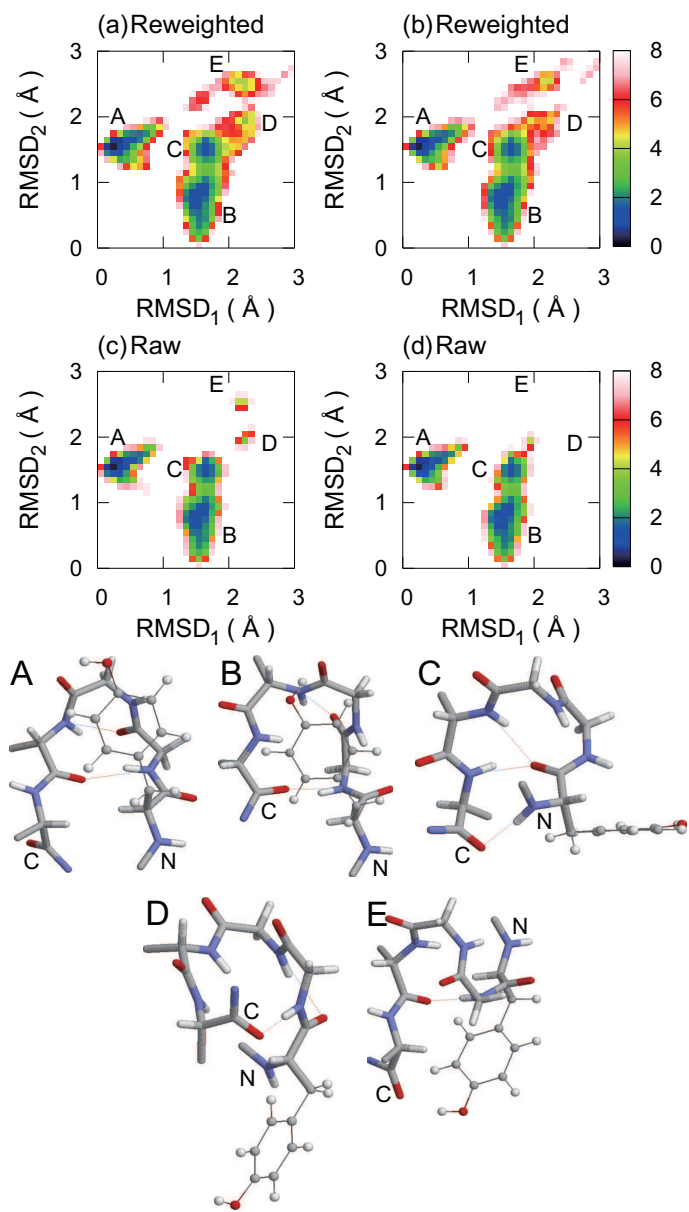}
\caption{
Free-energy landscapes at $T=200$~K obtained by the reweighting techniques 
from the (a) RPMD and (b) REMD simulations and 
those calculated from the raw histograms obtained by the
(c) RPMD and (d) REMD simulations.
The abscissa is the RMSD with respect to the structure in Fig~\ref{reference_enke:fig}(a).
The ordinate is the RMSD with respect to the structure in Fig~\ref{reference_enke:fig}(b).
The unit of the free-energy landscape is kcal/mol.
The labels A to E show the global-minimum and local-minimum free-energy states.
The representative conformations at these states are also presented.
The dotted lines denote hydrogen bonds.
The figures were drawn by RasMol \cite{rasmol}. 
}
\label{reweight_rmsd_prm:fig}
\end{figure}

Its local-minimum free-energy structures 
in vacuum have been reported for various force fields
although those for the AMBER parm99SB force field 
\cite{amber99sb} have yet to be reported.
For example, the global-minimum and local-minimum structures for
the CHARMM22 force field \cite{p22} are shown in Fig.~\ref{reference_enke:fig}.
To obtain global-minimum and local-minimum structures 
in the AMBER parm99SB force field and to see
the sampling efficiency of RPM, we illustrate free-energy landscapes at
200 K in Fig.~\ref{reweight_rmsd_prm:fig}.
The abscissa and ordinate are the root-mean-square deviation (RMSD)
with respect to the structure in Fig~\ref{reference_enke:fig}(a) (RMSD$_1$) and that 
with respect to the structure in Fig~\ref{reference_enke:fig}(b) (RMSD$_2$), 
respectively.
Here, the RMSD is defined by 
\begin{equation}
{\rm RMSD} = {\rm min} \left[\sqrt{\displaystyle \frac{1}{n} \sum_{j}
\left( \mbox{\boldmath $q$}_j-\mbox{\boldmath $q$}^0_j \right)^2} \right]~,
\label{def_rmsd}
\end{equation}
where $n$ is the number of the backbone atoms in Met-enkephalin,
$\{ \mbox{\boldmath $q$}^0_j \}$ are the coordinates in
the reference conformation, 
and the minimization is over the rigid translations and 
rigid rotations for the coordinates of
the conformation $\{ \mbox{\boldmath $q$}_j \}$ 
with respect to the center of geometry. 
The free-energy landscapes in
Figs.~\ref{reweight_rmsd_prm:fig}(a) and \ref{reweight_rmsd_prm:fig}(b)
were calculated from Eq.~(\ref{def_free}) with the reweighting techniques.
These landscapes show good agreement with each other,
and have five local-minimum free-energy states.
The five local-minimum states are labeled as A to E, as shown in the figures.
The free-energy landscapes in Figs.~\ref{reweight_rmsd_prm:fig}(c) and
\ref{reweight_rmsd_prm:fig}(d) were obtained from
the raw histogram without the reweighting techniques. 
In the landscape of REM in Fig.~\ref{reweight_rmsd_prm:fig}(d), 
States D and E are not observed although these states are observed in RPM.

In order to discuss sampling efficiency at the lowest temperature more quantitatively,
we counted the numbers of visiting times in each state, 
as listed in Table~\ref{visit_enke:table}.
Here, when a replica visited a local-minimum state at the lowest temperature
after the replica had visited another state at the lowest temperature,
it is counted as one visit.
The errors were estimated by the jackknife method \cite{quenouille56,miller74,bergbook} in
which the production run was divided into 20 segments.
We regarded the regions presented in Table~\ref{minima_enke:table} as
those for the local-minimum states.
As shown in Table~\ref{visit_enke:table},
REM did not visit State D and State E at the lowest temperature.
This is the reason why these states were not observed in Fig.~\ref{reweight_rmsd_prm:fig}(d).
On the other hand, RPM visited all states, and the numbers of visiting times in
RPM are larger than REM for all states.
RPM thus samples the conformational space more efficiently than
REM at the lowest temperature.

The representative conformations at
the local-minimum free-energy states are also shown 
in Fig.~\ref{reweight_rmsd_prm:fig}.
The structural features are as follows:
(State A) The structure in State A is the global-minimum free-energy structure for
the AMBER parm99SB force field and similar to that for
the CHARMM22 force field in Fig.~\ref{reference_enke:fig}(a).
This structure has two hydrogen bonds between NH of Gly2 and CO of Phe4 and
between CO of Gly2 and NH of Phe4.
The hydroxy group of the Tyr1 side-chain is close to CO of Gly3.
(State B) The structure in State B is almost the same as the structure in
Fig.~\ref{reference_enke:fig}(b).
Two hydrogen bonds are formed between NH of Gly2 and CO of Met5 and
between CO of Gly2 and NH of Phe4.
However, this structure does not have the hydrogen bond between
CO of Gly2 and NH of Met5 which exists in
Fig.~\ref{reference_enke:fig}(b).
Distance between the hydroxy group of Tyr1 and CO of Gly3 is small, as in State A.
(State C) There are two hydrogen bonds
between CO of Tyr1 and NH of Phe4 and
between CO of Tyr1 and NH of Met5 in State C.
(State D) Two hydrogen bonds between CO of Tyr1 and NH of Gly3 and
between NH of Gly2 and CO of Met5 are formed in State D.
As for the structures in States C and D, the Tyr1 hydroxy group is not close to
any backbone CO.
(State E) The structure in State E has a hydrogen bond between
NH of Gly2 and CO of Phe4.
This structure also has a small distance between the Tyr1 hydroxy group and CO of Met5.

  \subsection{C-peptide in Explicit Water}
  \label{cpep_rd:subsec}

\begin{figure}
\includegraphics[width=13cm,keepaspectratio]{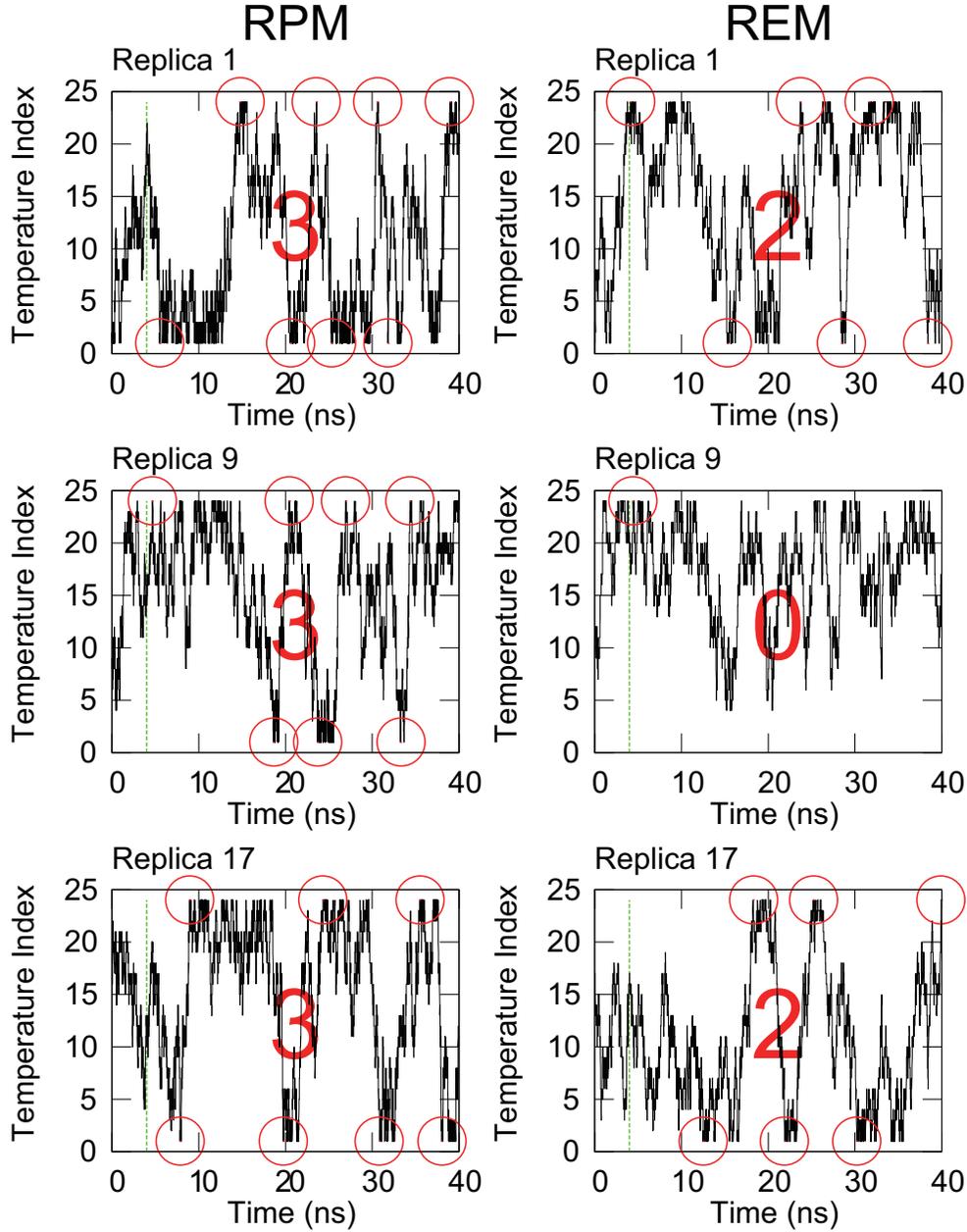}
\caption{
Time series of the temperatures of Replica 1, Replica 9, and Replica 17.
The left-hand figures and the right-hand figures are obtained from
the RPMD simulation and the REMD simulation, respectively.
The temperatures are represented as the temperature indices.
Index 1 and 24 correspond the lowest and highest temperatures, respectively. 
The production runs started from the green dashed lines.
Red circles indicate the steps at which the replicas reached
the highest (lowest) temperature after they had visited 
the lowest (highest) temperature.
The red numbers present the tunneling times of the replicas.
}
\label{time_prm:fig}
\end{figure}

The transition ratios of the replicas and the total numbers of tunneling times
for C-peptide in explicit water are shown in
Fig.~\ref{trans_ratio:fig}(c) and Table~\ref{tunnel:table}, respectively.
The transition ratios in RPM are larger than those in REM at all temperatures, again.
The tunneling times of RPM was about 2.1 times larger than that of REM.
The time series of the temperatures of Replica 1, Replica 9,
and Replica 17 are shown in Fig.~\ref{time_prm:fig}.
Figure~\ref{time_prm:fig} actually shows 
more frequent tunneling in RPM than in REM. 
In REM 6 replicas had never made a round trip
between the lowest and highest temperatures during the simulation.
In contrast, all replicas had at least one tunneling in RPM.
Most of them had more than two tunnelings.
Therefore, RPM samples the temperature space efficiently than REM.

\begin{figure}
\includegraphics[width=10cm,keepaspectratio]{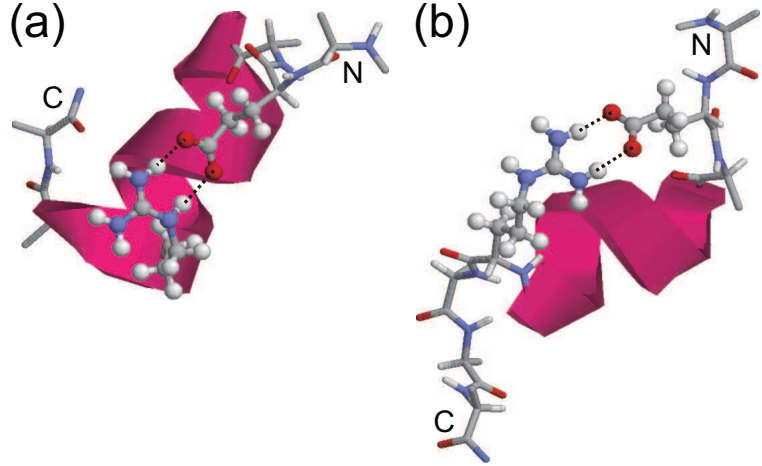}
\caption{
The lowest potential-energy conformations of C-peptide, 
which had $\alpha$-helix structures with SBs between Glu2 and Arg10,
obtained from the (a) RPMD and (b) REMD simulations.
The dotted lines denote the hydrogen bonds between the side-chains.
The figures were created with RasMol \cite{rasmol}. 
}
\label{cpeptide_helix:fig}
\end{figure}

It was reported in experiments that C-peptide has a helix structure with
salt bridges (SBs) between Glu2 and Arg10 at a low temperature \cite{obysdw89,fsysb90}.
We obtained helix structures which had such SBs in our RPMD and REMD simulations, 
as in these reports. 
The lowest potential-energy conformation among these helix structures for
each simulation is presented in Fig.~\ref{cpeptide_helix:fig}.
Here, we employed the DSSP (define secondary structure of proteins) 
criteria \cite{dssp} for
hydrogen bonds between the side-chains of Glu2 and Arg10 and 
for secondary structures of C-peptide.
The structure in Fig.~\ref{cpeptide_helix:fig}(a) 
obtained from the RPMD simulation 
has two hydrogen bonds between
${\rm O}_{\epsilon}$ of Glu2 and ${\rm H}_{\eta}$ of Arg10 and between
${\rm O}_{\epsilon}$ of Glu2 and ${\rm H}_{\epsilon}$ of Arg10.
The residues from Ala4 to Ala11 form the $\alpha$-helix structure.
As for the structure from the REMD simulation 
in Fig.~\ref{cpeptide_helix:fig}(b), 
the two ${\rm O}_{\epsilon}$ atoms of Glu2 form 
the hydrogen bonds with the two ${\rm H}_{\eta}$ atoms of Arg10.
The $\alpha$-helix structure is formed between Ala4 and Leu9.
Although these structures are slightly different with each other, 
both RPM and REM sampled conformations near
the lowest potential-energy helix structure in the other method.

\begin{figure}
\includegraphics[width=16cm,keepaspectratio]{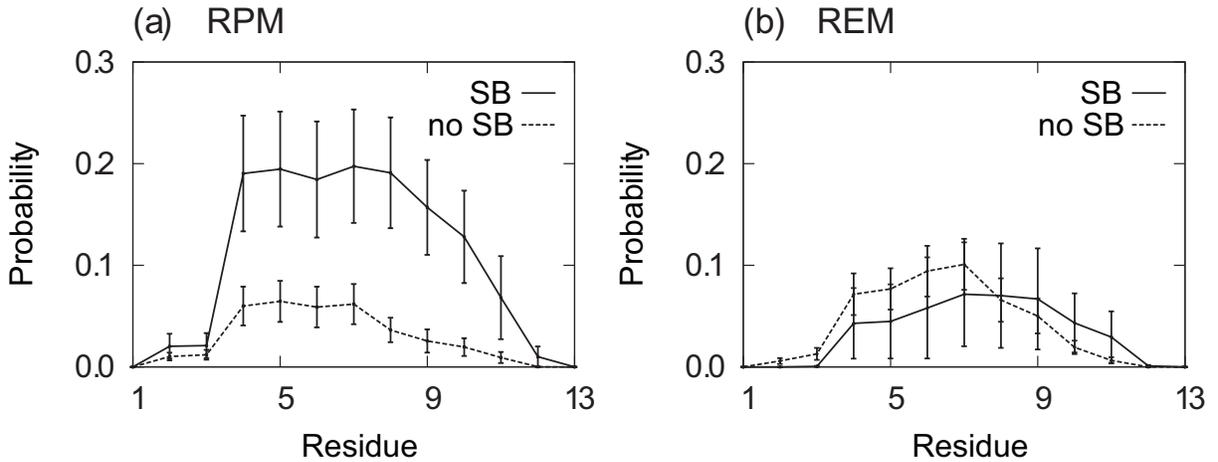}
\caption{
Probabilities of $\alpha$-helix structures at 281 K for each residue in
the (a) RPMD and (b) REMD simulations.
Solid line and dashed line show the results with and without SBs
between Glu2 and Arg10, respectively.
}
\label{reweight_prob_salt_jack:fig}
\end{figure}

To see effects of the SBs on the $\alpha$-helix structures,
we calculated probabilities of the $\alpha$-helix structures with the SBs
as well as without them.
These probabilities at 281 K for each residue are shown in
Fig.~\ref{reweight_prob_salt_jack:fig}.
The probabilities without the SBs 
in RPM agree well with those in REM.
The probabilities for residues 4 to 7 in both methods are high
regardless of the existence of the SBs.
This is because their amino-acid sequence is AAAK, and
this sequence is known for having $\alpha$-helix structures \cite{al05}.
In RPM the probabilities with the SBs are especially higher than those without
the SBs while both probabilities are almost the same in REM.
We discuss the origin of this difference between RPM and REM later.

\begin{figure}
\includegraphics[width=16cm,keepaspectratio]{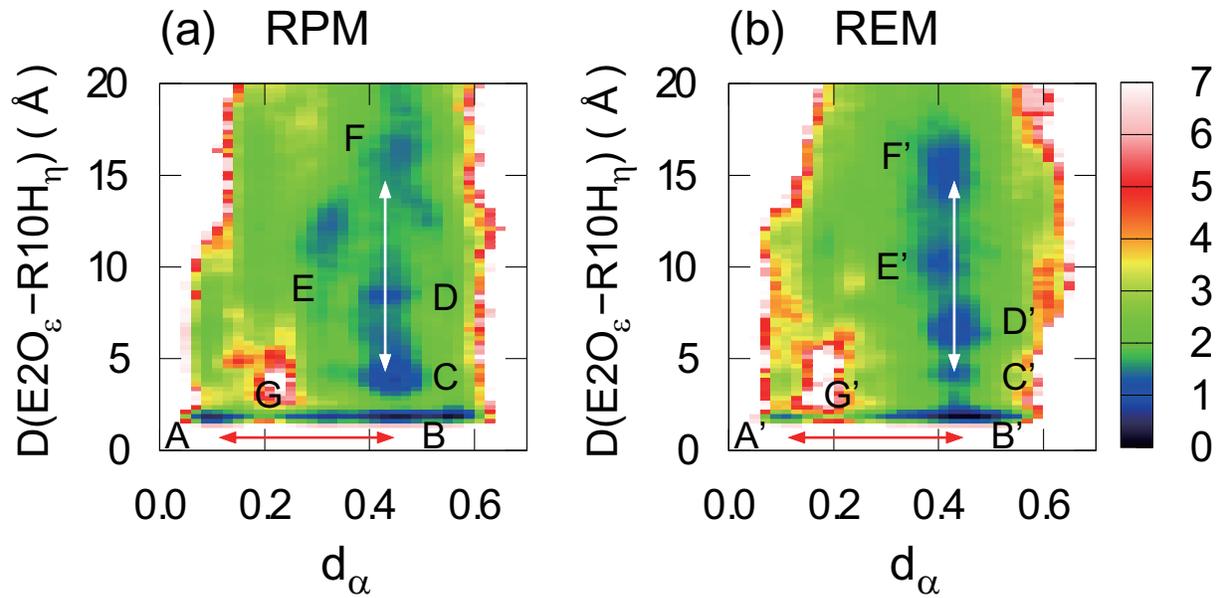}
\caption{
Free-energy landscapes at 281 K obtained from the (a) RPMD and (b) REMD simulations.
The abscissa is the dihedral-angle distance with respect to
the reference $\alpha$-helix structure.
The ordinate is distance between 
${\rm O}_{\epsilon}$ of Glu2 and ${\rm H}_{\eta}$ of Arg10.
The unit of the free-energy landscapes is kcal/mol.
The labels A to F and A$^{\prime}$ to F$^{\prime}$ show the local-minimum free-energy states.
The labels G and G$^{\prime}$ are the transition states between
States A and B and between States A$^{\prime}$ and B$^{\prime}$, respectively.
}
\label{reweight_free_dist_ddista:fig}
\end{figure}

It is considered that the SBs between Glu2 and Arg10 stabilize
the $\alpha$-helix structure of C-peptide \cite{fsysb90}.
In order to investigate the relation between the SBs and
the stability of the $\alpha$-helix structure,
we calculated a free-energy landscape at 281 K in each method from
Eq.~(\ref{def_free}) with the reweighting techniques.
These free-energy landscapes are shown in Fig.~\ref{reweight_free_dist_ddista:fig}.
The abscissa is the dihedral-angle distance $d_{\alpha}$ with respect to
a reference $\alpha$-helix structure.
Here, a dihedral-angle distance $d_{\alpha}$ is defined by
\begin{equation}
d_{\alpha} = \frac{1}{n \pi} \sum_{i=1}^{n} \delta (v_{i},v_{i}^{0})~,
\label{def_dd}
\end{equation}
where $n$ is the total number of dihedral angles, 
$v_{i}$ is the dihedral angle $i$,
and $v_{i}^{0}$ is the dihedral angle $i$ of the reference conformation. 
The distance $\delta (v_{i},v_{i}^{0})$ between two dihedral angles is given by 
\begin{equation}
\delta (v_{i},v_{i}^{0}) = {\rm min} (|v_{i} - v_{i}^{0}|,2 \pi - |v_{i} - v_{i}^{0}|)~.
\label{def_dis}
\end{equation}
For $d_{\alpha}$, only the backbone-dihedral angles of
the residues 4-10 were employed as the elements in Eq.~(\ref{def_dd}).
We set the value of $v_{i}^{0}$ to $(\phi,\psi)=(-\pi/3,-\pi/3)$.
When the value of $d_{\alpha}$ is close to 0, therefore,
C-peptide has a $\alpha$-helix structure.
The ordinate is distance between 
${\rm O}_{\epsilon}$ of Glu2 and ${\rm H}_{\eta}$ of Arg10: 
$D({\rm E}2{\rm O}_{\epsilon}-{\rm R}10{\rm H}_{\eta})$~.
Here, this distance is defined to be the smallest 
among the four sets of the distances between
two ${\rm O}_{\epsilon}$ atoms of Glu2 and two ${\rm H}_{\eta}$ atoms of Arg10.
Six local-minimum free-energy states are observed in both RPM and REM as shown in
Figs.~\ref{reweight_free_dist_ddista:fig}(a) and (b).
We label these local-minimum states as A to F for RPM and
A$^\prime$ to F$^\prime$ for REM.
The transition states between
States A and B and between States A$^{\prime}$ and B$^{\prime}$,
are labelled as G and G$^{\prime}$, respectively.
The $\alpha$-helix structure with the SBs as in
Fig.~\ref{cpeptide_helix:fig} corresponds to a structure in State A or A$^\prime$.
This fact means that the $\alpha$-helix structure with the SBs is a stable structure.
On the other hand,
the $\alpha$-helix structure without the SBs is not a stable structure
because there is no local-minimum states for this structure.
Therefore, the SBs play an important role in stabilizing
the $\alpha$-helix structure.

The global-minimum state at 281 K in RPM is State A while
that in REM is State B$^\prime$.
To see the reason for this difference,
we counted the number of visiting times in State A for RPM and
in State A$^\prime$ for REM at the lowest temperature during the simulations.
The region from 0.00 to 0.17 for $d_{\alpha}$ and 
from 1.0 \AA~to 2.2 \AA~for 
$D({\rm E}2{\rm O}_{\epsilon}-{\rm R}10{\rm H}_{\eta})$
was assigned to State A and State A$^\prime$ here. 
The number of visiting times for each method is listed 
in Table~\ref{visit_cpeptide:table}.
Here, we employed two criteria to count the number of visiting times.
In Criterion 1, when a replica visited in State A (or A$^\prime$) at
the lowest temperature after the replica had sampled $d_{\alpha}$ 
larger than 0.25 or 
$D({\rm E}2{\rm O}_{\epsilon}-{\rm R}10{\rm H}_{\eta})$ 
larger than 3.2 \AA~at the lowest temperature,
it is counted as one visit.
In Criterion 2, 
sampling $D({\rm E}2{\rm O}_{\epsilon}-{\rm R}10{\rm H}_{\eta})$ 
larger than 3.2 \AA~was not taken into account.
Therefore, the number of visiting times increases only when
a conformation is changed from a non-helical structure to a helix structure
with the SBs in Criterion 2.
In addition to this,
breaking and forming of
the SBs also increases the number of visiting times in Criterion 1.
As shown in Table~\ref{visit_cpeptide:table},
the numbers of visiting times in RPM are much larger 
than those in REM in both criteria.
In Criterion 2, especially, the number of visiting times is 1 in REM.
Because of such insufficient sampling in State A$^\prime$,
this state was underestimated in REM.
This underestimation caused the small probabilities of
the $\alpha$-helix structures with the SBs in 
Fig.~\ref{reweight_prob_salt_jack:fig}(b).

\begin{figure}
\includegraphics[width=13cm,keepaspectratio]{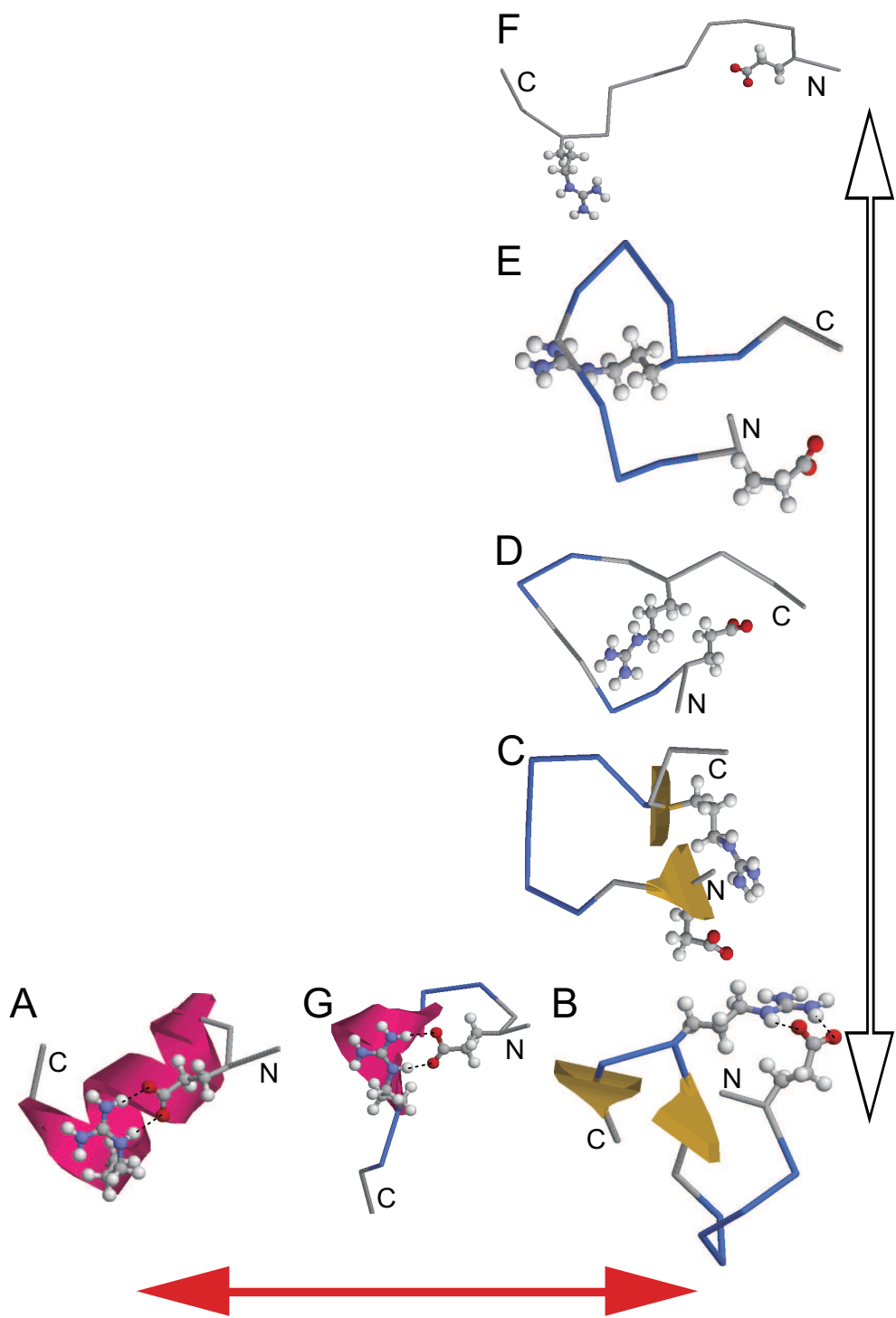}
\caption{
The representative conformations of the local-minimum free-energy states A to F 
and the transition state G obtained by the RPMD simulation.
The dotted lines denote the SBs between Glu2 and Arg10.
Backbone blue color shows turn structures.
The figures were drawn by RasMol \cite{rasmol}. 
}
\label{structure_cpep:fig}
\end{figure}

The representative conformations for each state in
Fig.~\ref{reweight_free_dist_ddista:fig}(a) are shown 
in Fig.~\ref{structure_cpep:fig}.
The representative conformation for the transition state G is also presented.
From these conformations and the free-energy landscape,
we clarify a folding pathway from an extended structure in State F 
to the $\alpha$-helix structure with the SBs in State A.
Because the pathways in RPM and REM are almost the same,
we will discuss that only in RPM:
(State F to State C via States E and D)
The extended structure of C-peptide in State F changes to
a globular structure as the side-chains of Gly2 and Arg10 
get close together.
Turn structures are formed between Gly2 and Arg10 by this conformational change,
as in States D and E.
In State C the antiparallel $\beta$-bridge structure is occasionally created between
Gly2 and Arg10.
(State C to State B) 
The SBs are formed between Gly2 and Arg10 by coming close together.
Antiparallel $\beta$-bridges between Thr3 and Lys7 or between
Leu9 and His12 are occasionally observed in State B.
(State B to State A via State G)
In State G, a short $\alpha$-helix or 3$_{10}$-helix structure
is formed around residues 7 to 10, while maintaining the SBs.
By growing this helix structure,
the longer $\alpha$-helix structure with the SBs
is formed between residues 4 and 11 as in State A.

This folding pathway can be divided into two steps.
The first step is the ``salt-bridge formation step'',
and the second step is the ``$\alpha$-helix formation step''.
The first step and second step correspond to
the transition from State F to B and that from B to A, respectively.
These steps are drawn by the white and red arrows in
Figs.~\ref{reweight_free_dist_ddista:fig} and \ref{structure_cpep:fig}.
We can also see in Fig.~\ref{reweight_free_dist_ddista:fig} that 
C-peptide rarely takes a folding pathway
in which the salt-bridge is formed after
the $\alpha$-helix structure formation. 

  \section{Conclusions} 
  \label{conc:sec}

We proposed the replica-permutation method (RPM), in which
the replicas are allowed to transit not only neighboring temperatures
but also non-neighboring temperatures.
For replica-permutation trials in this method,
the Suwa-Todo algorithm was employed instead of the Metropolis algorithm.
This is because 
most of the permutation trials are rejected in the Metropolis algorithm.
The Suwa-Todo algorithm had been proposed originally 
to minimize average rejection ratios for
state transitions in MCMC.
We applied RPM and M-RPM to the particles with the double-well potential energy to
clarify the usefulness of the Suwa-Todo algorithm for the replica permutations.
For a comparison purposes,
REMD simulations were also performed.
As a result, RPM realized the most efficient sampling in the temperature space
while replica permutations were hardly accepted in M-RPM.

We also applied RPM and REM to Met-enkephalin in vacuum.
RPM sampled the temperature space more efficiently than REM
even in the biomolecular system.
The five local-minimum free-energy states
were obtained at 200 K in both methods by using the reweighting techniques.
In the free-energy landscape estimated from
the raw histogram in REM, however,
two of the five local-minimum states were not observed.
This is because REM did not sample these two states at the lowest temperature.
On the other hand, RPM sampled all states even at the lowest temperature.
It indicates that RPM realized efficient sampling not only in
the temperature space but also in the conformational space.

Furthermore, the RPMD and REMD simulations were performed
for the C-peptide analog in explicit water to
see the usability of RPM for a larger biomolecular system.
RPM showed higher sampling efficiency in the temperature space, again.

It is reported in experiments that C-peptide has
the $\alpha$-helix structure with the SBs between Gly2 and Arg10.
We observed the $\alpha$-helix structures in both simulations.
We also showed that the SBs play an important role in stabilizing
the $\alpha$-helix structure.
From the free-energy landscape, furthermore,
the folding pathway from the extended structure to
the $\alpha$-helix structure with the SBs was clarified.
This folding pathway consists of the two steps.
The first step is the ``salt-bridge formation step''. 
In this step, the SBs are formed by changing 
its conformation from the extended structure to the globular structures.
The second step is the ``$\alpha$-helix formation step''.
The $\alpha$-helix structure is formed while maintaining the SBs.

We thus revealed that RPM realizes more efficient sampling 
in the conformational space at
the low temperature than REM.
Furthermore, because the transition ratios of the replicas 
in RPM were larger than those in REM at all temperatures for all systems,
larger temperature intervals can be taken in RPM.
Therefore, the number of replicas can be reduced.

Although only the results of the MD simulations of RPM were shown in this article,
this method can be readily applied to the MC method.
It is also straightforward to introduce this method to
the multidimensional REM \cite{sko00} (also called Hamiltonian REM \cite{fwt02}) and
related methods \cite{ioo10}.
We can enhance the sampling efficiency of these methods by replacing REM to RPM.

\section*{ACKNOWLEDGMENTS}

The computations were performed on the computers at 
the Research Center for Computational Science, Okazaki Research Facilities,
National Institutes of Natural Sciences.


\clearpage

%
\begin{table}
\caption{Example of an assignment of the labels in RPM with 3 replicas.
}
\label{ex_state:table}
\vspace{0.5cm}
\begin{tabular}{c|c|c|c|c|c|c}
\hline \hline
\backslashbox{Temperature}{Label $\alpha$ of $X_{\alpha}$} & 1 & 2 & 3 & 4 & 5 & 6 \\ 
\hline
$T_1$ & Replica 1 & Replica 1 & Replica 2 & Replica 2 & Replica 3 & Replica 3 \\
$T_2$ & Replica 2 & Replica 3 & Replica 1 & Replica 3 & Replica 1 & Replica 2 \\
$T_3$ & Replica 3 & Replica 2 & Replica 3 & Replica 1 & Replica 2 & Replica 1 \\
\hline \hline

\end{tabular}

\vspace{0.5cm}
\end{table}
\begin{table}
\caption{The total numbers of tunneling times during the simulations.
}
\label{tunnel:table}
\vspace{0.5cm}
\begin{tabular}{cccc}
\hline \hline
 & Double-Well$^{*}$ & Enkephalin & C-peptide \\ 
\hline
RPM   & 424 $\pm$ 64 & 459 $\pm$ 21 & 58 $\pm$ 6 \\
REM   & 254 $\pm$ 39 & 357 $\pm$ 14 & 27 $\pm$ 4 \\
M-RPM &  14 $\pm$  3 &              &            \\
\hline \hline

\end{tabular}
\vspace{0.3cm}

$^{*}$ The values were obtained by taking an average of the 40 simulations' results

\vspace{0.5cm}
\end{table}
\begin{table}
\caption{The number of visiting times in each state of Met-enkephalin in vacuum.
}
\label{visit_enke:table}
\vspace{0.5cm}
\begin{tabular}{cccccc}
\hline \hline
Method & A & B & C & D & E \\
\hline
RPM & 97 $\pm$ 9 & 98 $\pm$ 6 & 88 $\pm$ 7 & 1 $\pm$ 1 & 2 $\pm$ 1 \\
REM & 86 $\pm$ 5 & 86 $\pm$ 6 & 65 $\pm$ 5 & 0 $\pm$ 0 & 0 $\pm$ 0 \\
\hline \hline

\end{tabular}

\vspace{0.5cm}
\end{table}
\begin{table}
\caption{RMSD ranges for each local-minimum free-energy state.
}
\label{minima_enke:table}
\vspace{0.5cm}
\begin{tabular}{ccc}
\hline \hline
State & RMSD$_1$ (\AA) & RMSD$_2$ (\AA) \\
\hline
A & 0.0 - 0.6 & 1.3 - 1.8 \\
B & 1.3 - 1.7 & 0.4 - 1.0 \\
C & 1.5 - 1.8 & 1.3 - 1.7 \\
D & 2.1 - 2.4 & 1.8 - 2.1 \\
E & 2.0 - 2.4 & 2.3 - 2.8 \\
\hline \hline

\end{tabular}

\vspace{0.5cm}
\end{table}
\begin{table}
\caption{The numbers of visiting times in State A for RPM and in State A$^\prime$ for REM.
}
\label{visit_cpeptide:table}
\vspace{0.5cm}
\begin{tabular}{ccc}
\hline \hline
Method & Criterion 1 & Criterion 2 \\
\hline
RPM & 14 $\pm$ 8 & 5 $\pm$ 1 \\
REM &  4 $\pm$ 2 & 1 $\pm$ 0 \\
\hline \hline

\end{tabular}

\vspace{0.5cm}
\end{table}


\clearpage

%
%
%
%
%
%
%
%
%
%
%
%
%
%
%
%
%
%
%
%
%
%
%
\end{document}